\documentclass{appolb}
\usepackage{graphicx}

\begin{document}
\title{Multiplicity dependence of strangeness and charged particle production in proton-proton collisions %
\thanks{Presented at Epiphany 2019}%
}
\author{Prabhakar Palni (for the ALICE Collaboration)
\address{AGH University of Science and Technology}
}
\maketitle
\begin{abstract}
 In this contribution, the production rates and the transverse momentum distributions of strange
hadrons are reported as a function of charged particle multiplicity. In this analysis, the data collected in proton-proton collisions at  $\sqrt{s}$ = 13 TeV with the ALICE detector at the LHC are used. It is found that the production rate of $K_{S}^{0}$, $\Lambda$, $\Xi^{\pm}$, and $\Omega$ increases with
multiplicity faster than that for charged particles. The higher the strangeness
content of the hadron, the more pronounced is the increase. Moreover, the energy
and multiplicity dependence of charged particle production in pp collisions
are presented and the results are compared to  predictions from Monte Carlo (MC) event generators. It turns out that the average multiplicity density increases steeply with center-of-mass energy for high
multiplicity classes.

\end{abstract}
  
\section{Introduction}
A recent multiplicity dependent study of the strange and multi-strange hadron production in pp collisions has provided quite interesting results hinting at a Quark-Gluon Plasma (QGP) like properties in pp collisions \cite{nature}. The enhancement of strangeness production in high-energy nucleus-nucleus collisions has historically been proposed as one of the first signatures of QGP formation \cite{strange}. This work provides a comprehensive study of  energy and multiplicity dependence of charged-particle production measured with forward and central multiplicity estimators, which serve as an important reference for multiplicity dependent measurements in pp collisions. The multiplicity dependence of strange hadron production rates and their transverse momentum distributions  at center-of-mass energy 13 TeV are measured and compared with the previous 7 TeV results in this work.

\section{Charged particle production}
The description of the ALICE detector setup can be found in \cite{alice1,dndeta1}. This measurement relies on track segments called Silicon Pixel Detector (SPD) tracklets  covering the kinematic
region $|\eta| < 1.8$.  

\begin{figure}[h]
	\centering
	\includegraphics[height=0.6\textheight,width=0.4\textwidth,keepaspectratio]{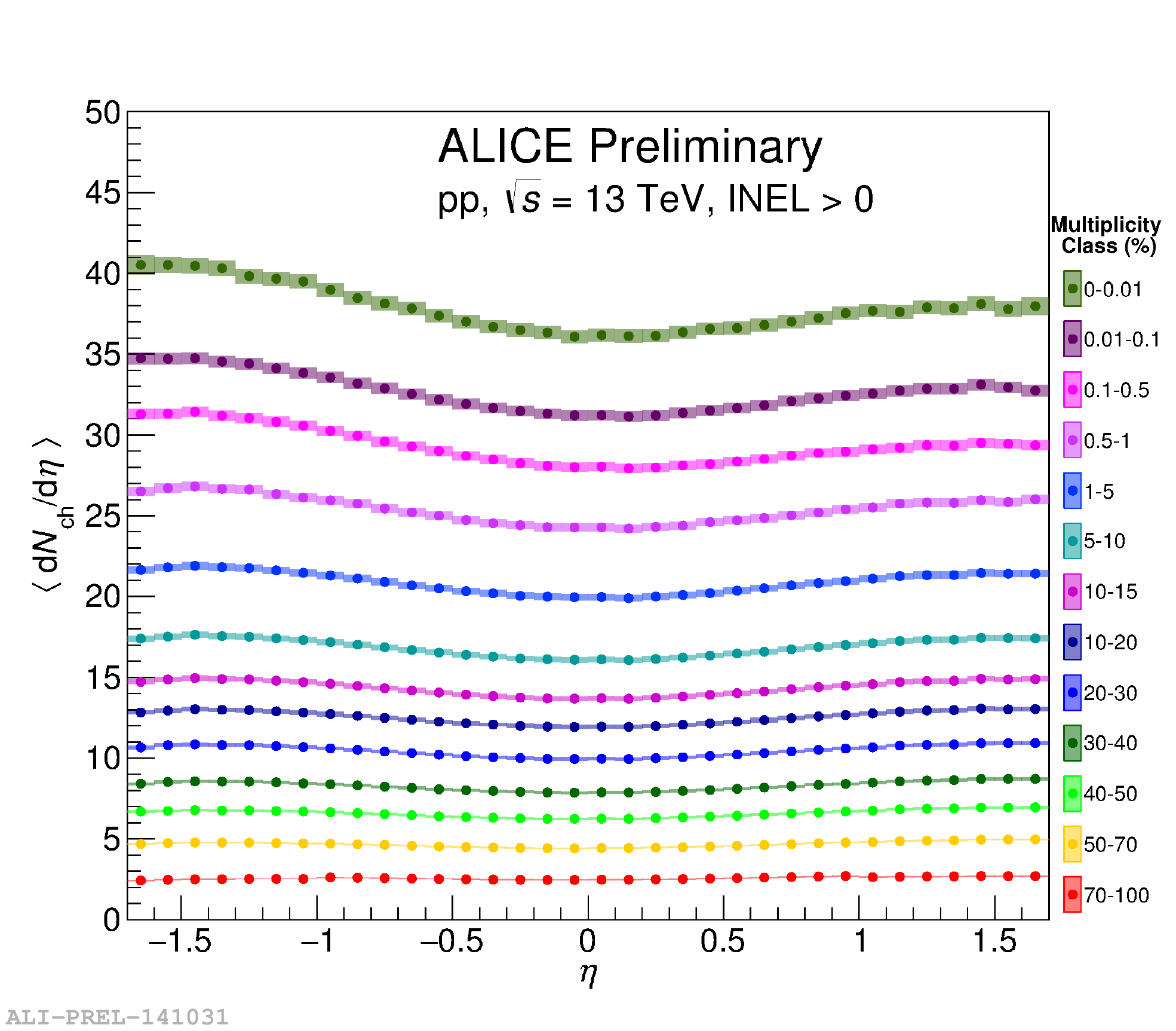}
	\includegraphics[height=0.6\textheight,width=0.45\textwidth,keepaspectratio]{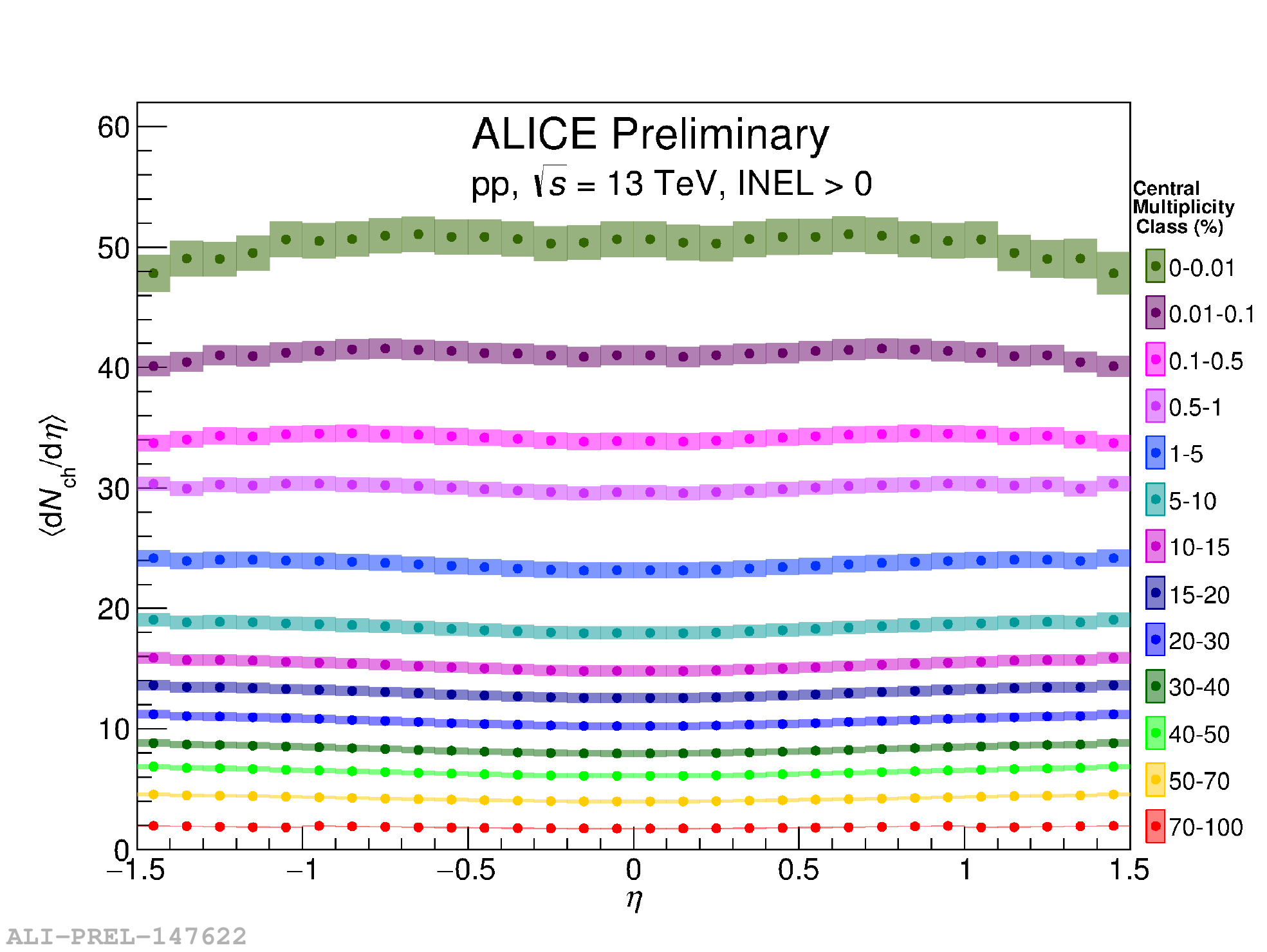}
	
	\caption{Multiplicity dependence of the pseudorapidity density distributions measured with forward (left) and central (right) multiplicity estimators.}\label{dndeta_forward}
\end{figure}

Results are presented for inelastic collisions with at least one charged particle in the central region $|\eta| < 1.0 $ (called the  ``INEL $> 0$" event class). Fig.~\ref{dndeta_forward}  shows the  charged-particle pseudorapidity-density distribution measured using a forward multiplicity estimator (total charge deposited in V0A and V0C scintillators) and a central multiplicity estimator (tracklets in the central SPD detector) for different multiplicity classes. These multiplicity classes are made by dividing the event samples based on the total charge deposited in V0A and V0C scintillators (for forward multiplicity estimator) and number of SPD tracklets  (for central multiplicity estimator) which is proportional to the charged particle production.
\begin{figure}[h]
	
	\includegraphics[height=0.5\textheight,width=0.40\textwidth,keepaspectratio]{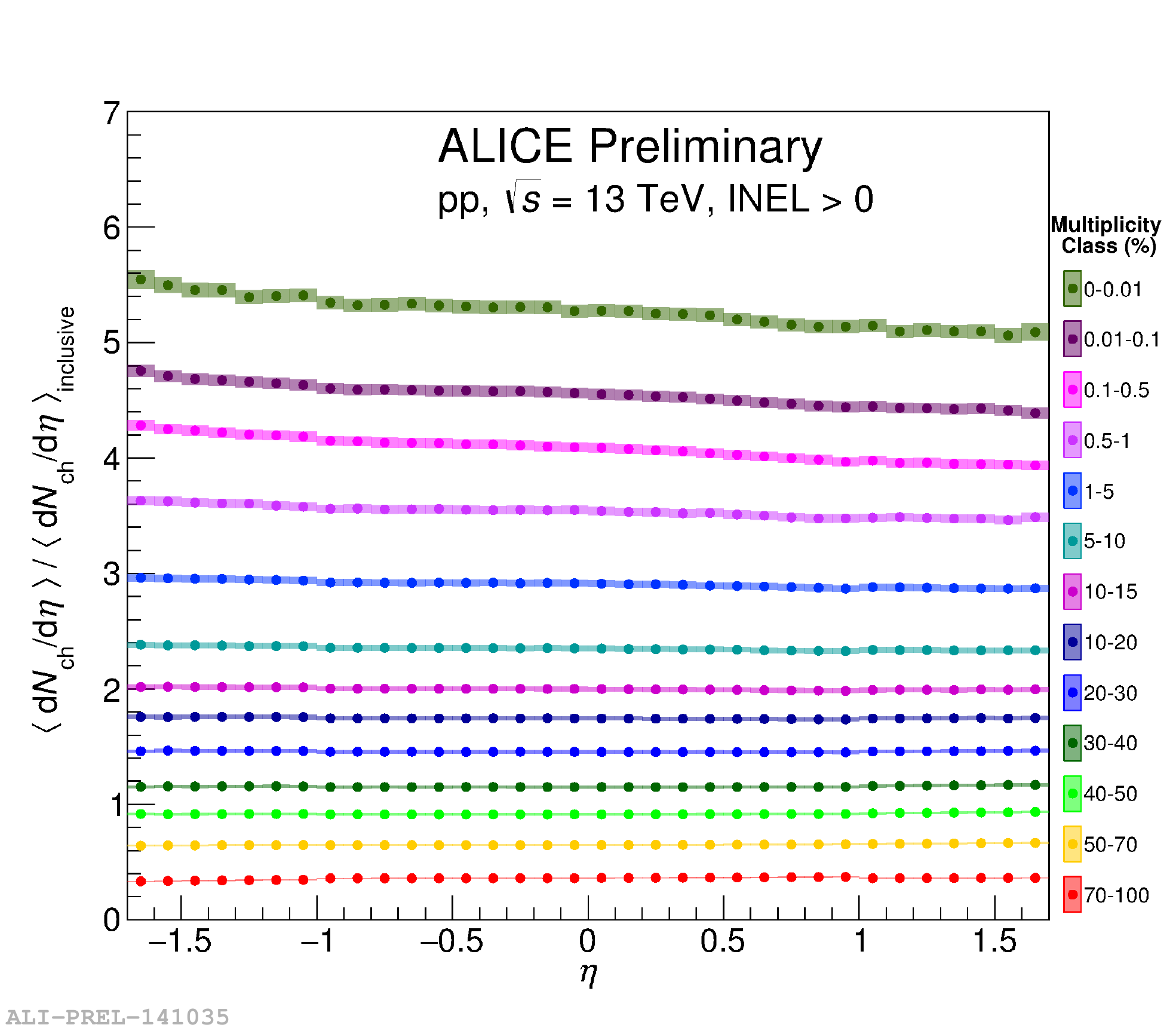}
	\includegraphics[height=0.5\textheight,width=0.46\textwidth,keepaspectratio]{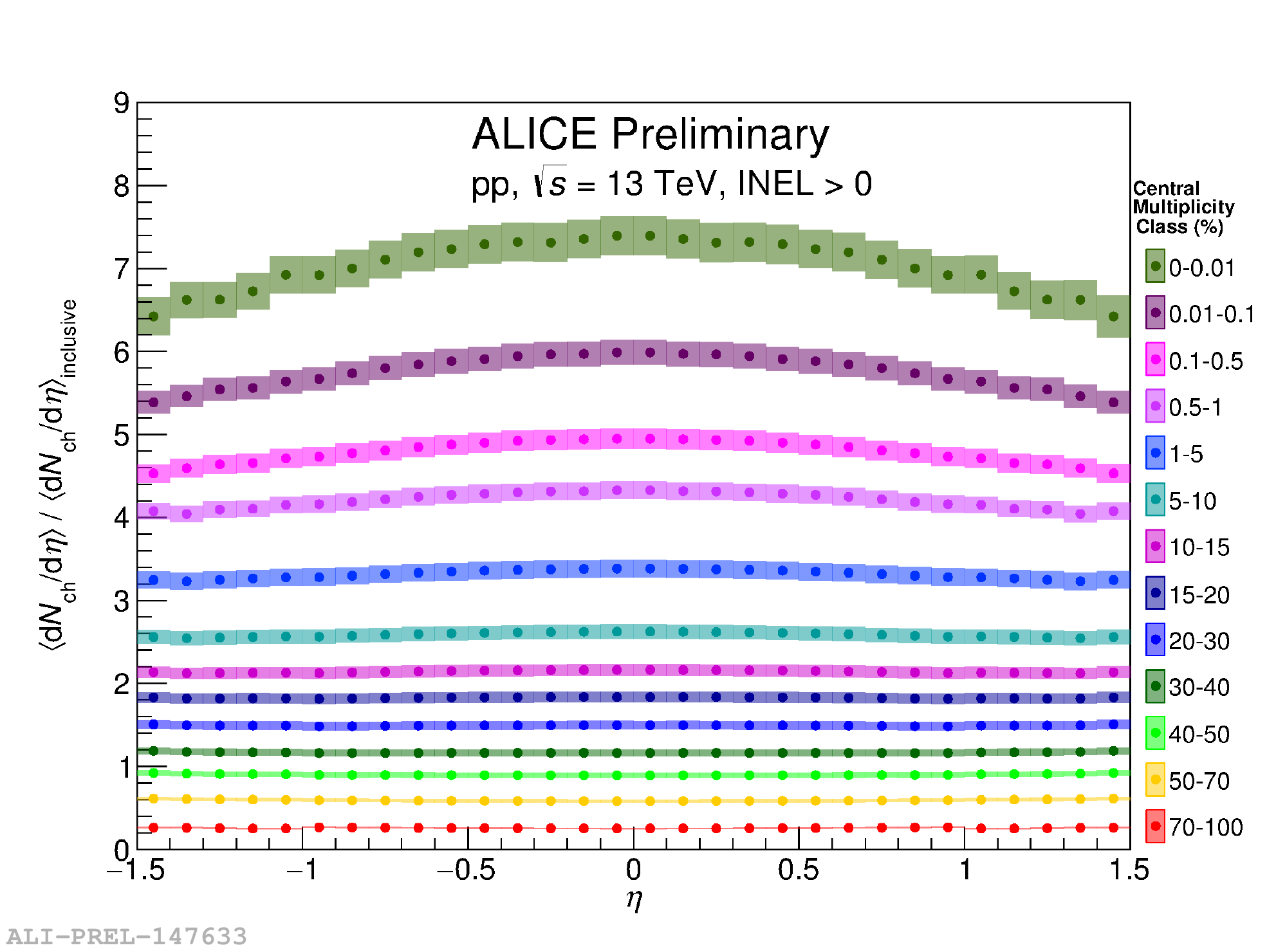}
	
	\caption{Normalized multiplicity dependence of the pseudorapidity density distributions measured with forward (left) and central (right) multiplicity estimators.}\label{norm_dndeta_forward}
\end{figure}
The observed asymmetry for the forward estimator reflects the asymmetric acceptances of our scintillator detectors V0A (which covers the pseudo-rapidity range $2.8 < \eta < 5.1$) and V0C (which covers the pseudo-rapidity range $-3.7 < \eta < 1.7$), which can also be seen in simulated data at the particle level. An advantage of using a forward multiplicity estimator is to  provide an unbiased measurement of $\langle \mathrm{d}\it{N}_{\mathrm{ch}}/\mathrm{d}\it{\eta} \rangle $ at mid-rapidity due to auto-correlation.

\begin{figure}[h]
	\centering
	
	\includegraphics[height=0.45\textheight,width=0.39\textwidth,keepaspectratio]{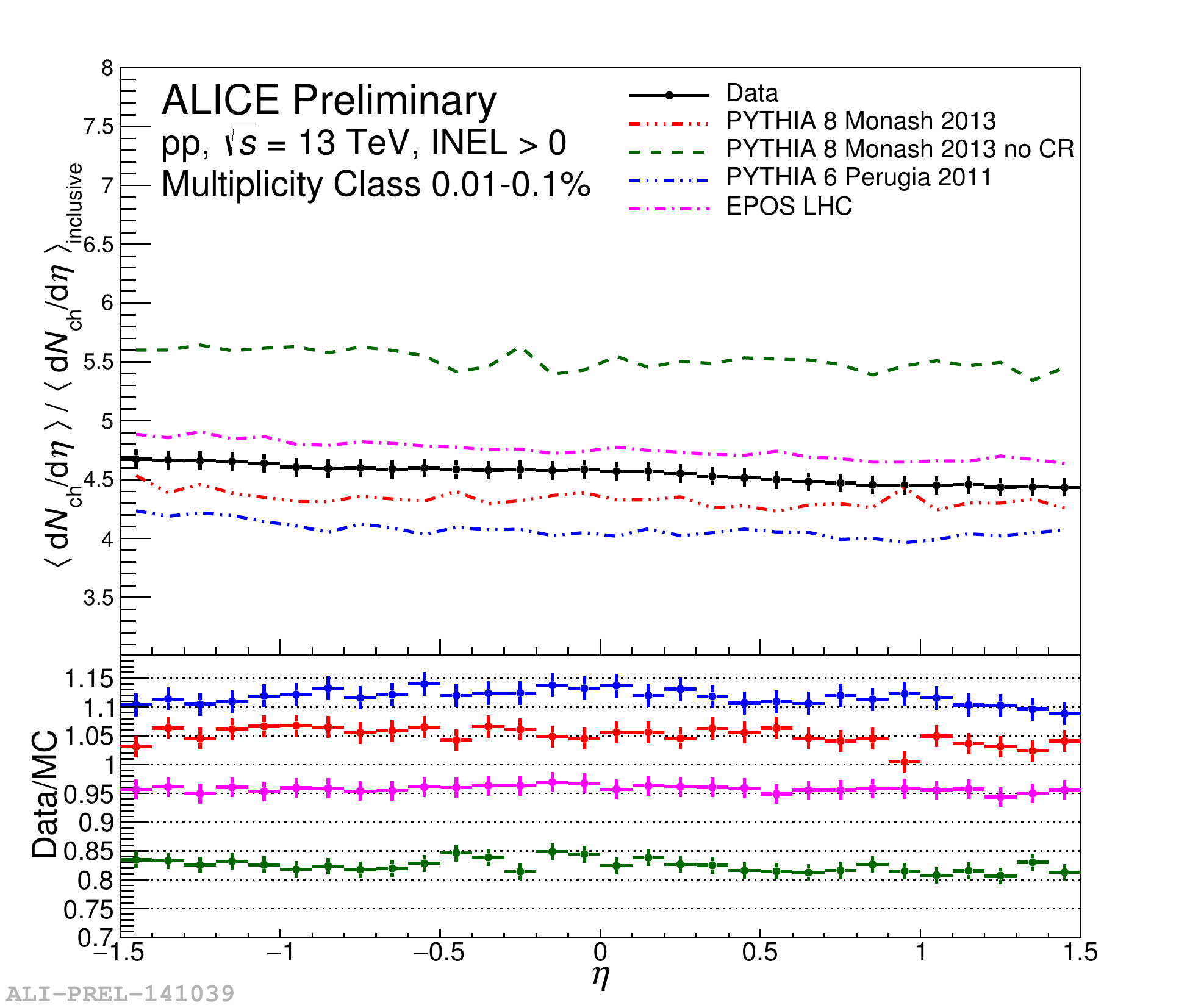}
	\includegraphics[height=0.52\textheight,width=0.47\textwidth,keepaspectratio]{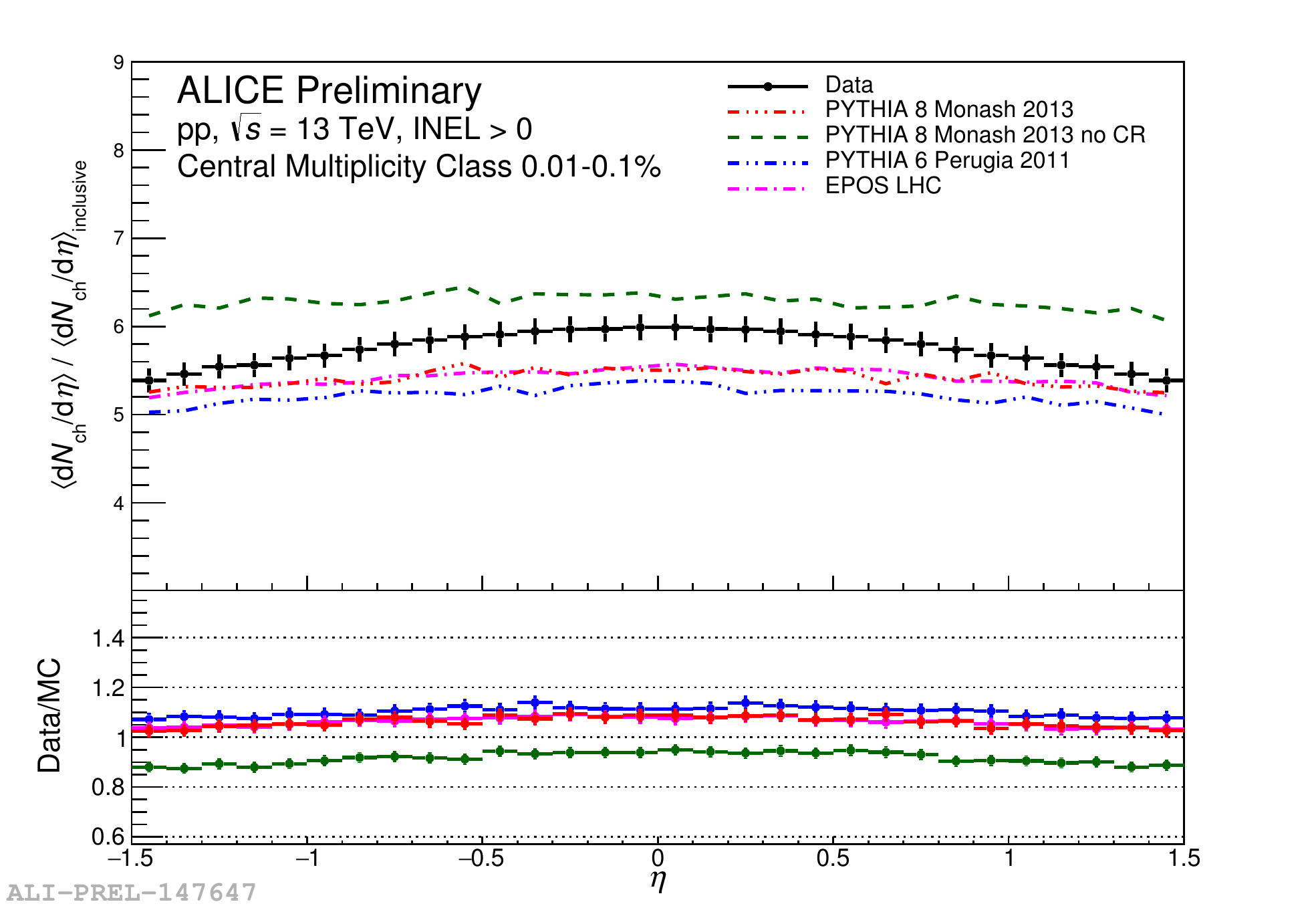}
	
	\caption{Normalized pseudorapidity density distribution of charged particles is compared with different MC models for multiplicity bin 0.01-0.1\%  for  forward (left) and central (right) multiplicity estimators.}\label{norm_dndeta00101_forward}
\end{figure}
Normalized pseudorapidity-density distributions are presented in Fig. \ref{norm_dndeta_forward}  for both forward  and central multiplicity estimators. We used the inclusive  pseudorapidity density distribution of charged particles to normalize the distributions. Using the normalized distribution we reduced most of the correlated systematic uncertainties. The normalized results  in Fig. \ref{norm_dndeta_forward} show up to $\sim$7 and  $\sim$5 times more charged particle production in the highest multiplicity class (0-0.01\%) with the central and forward estimators, respectively.

\begin{figure}[h]
	
	\includegraphics[height=0.48\textheight,width=0.44\textwidth,keepaspectratio]{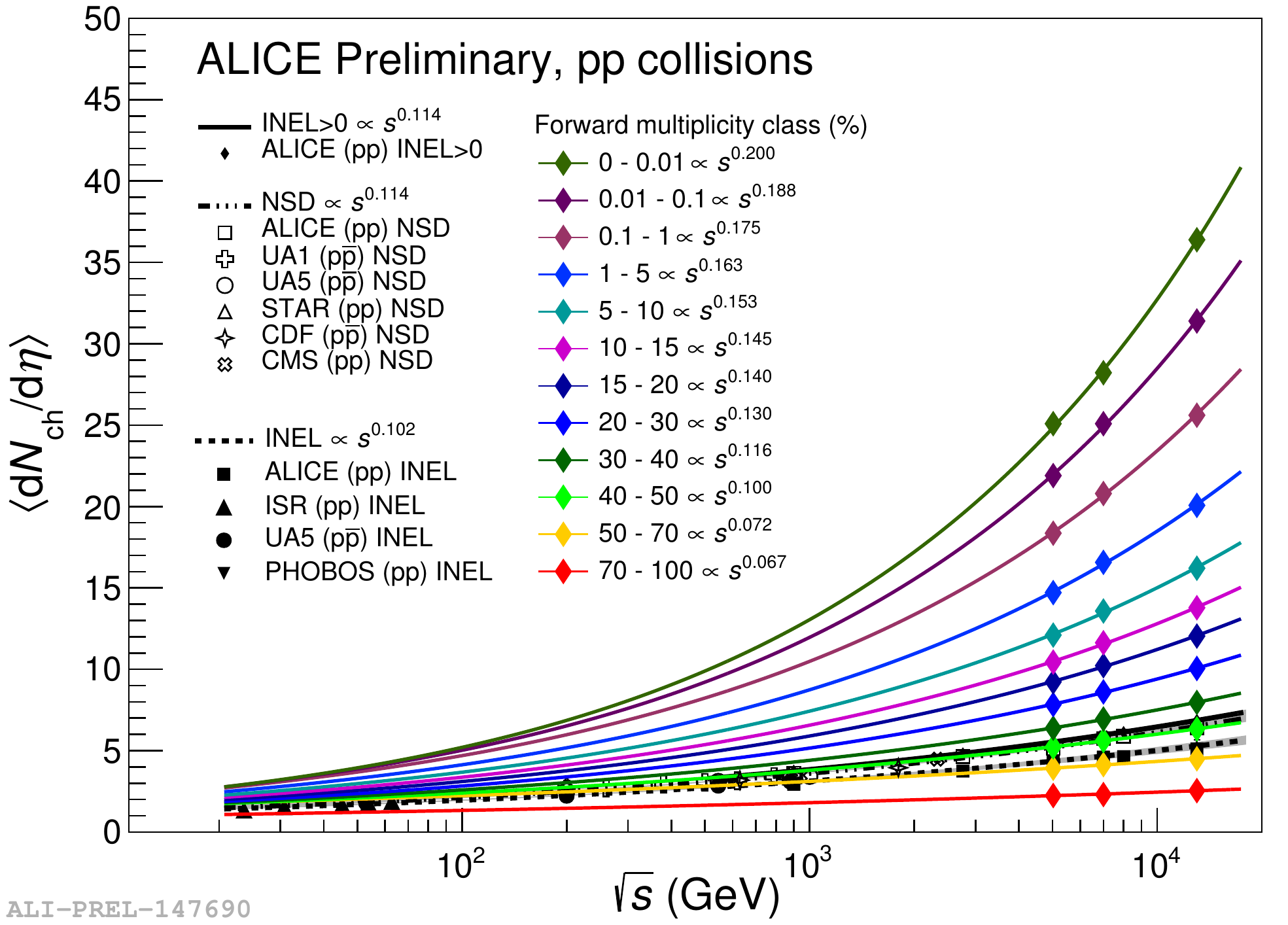}
	\includegraphics[height=0.42\textheight,width=0.44\textwidth,keepaspectratio]{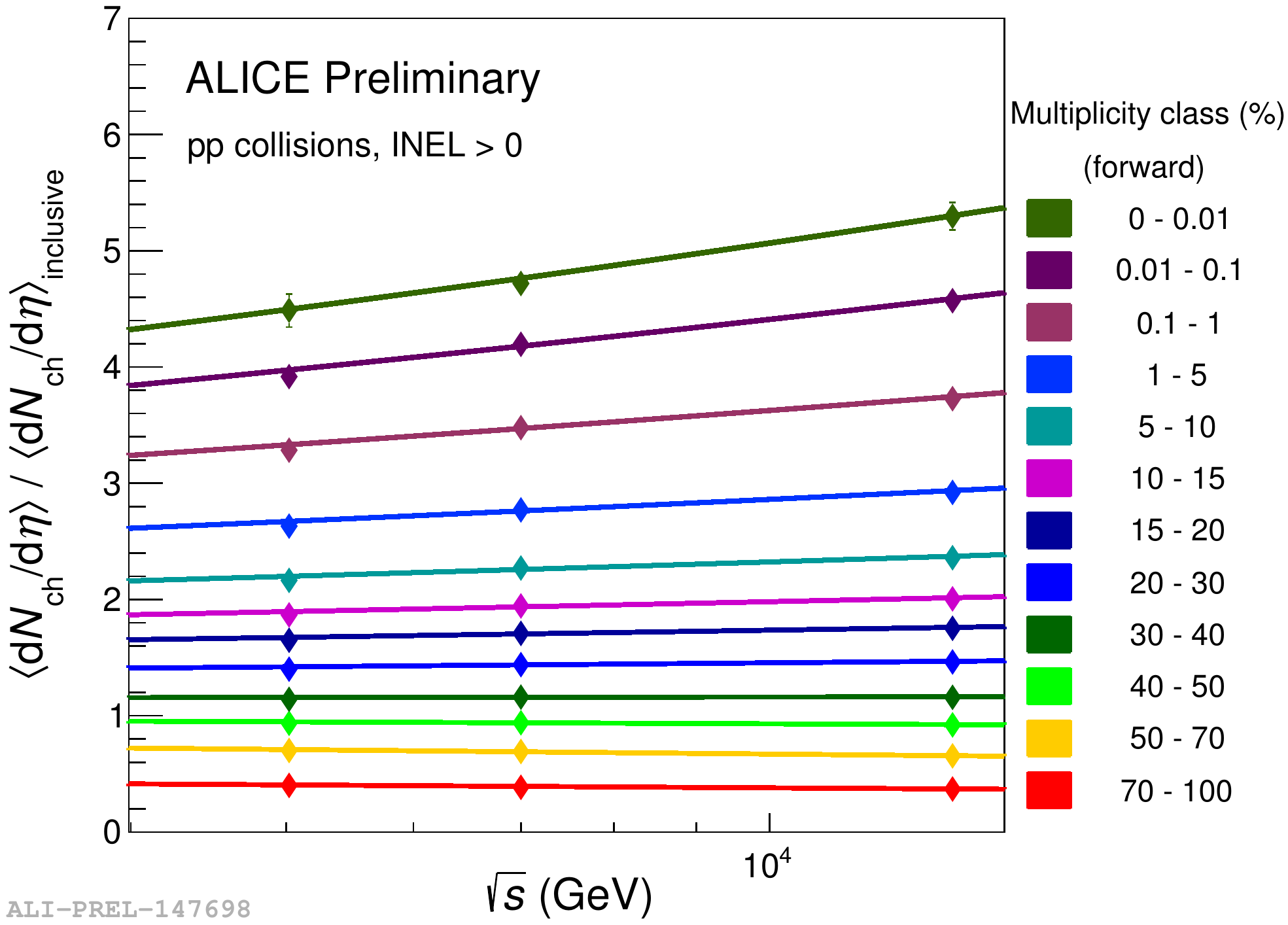}	
	\caption{Energy dependence of the pseudorapidity density distributions in pp collisions at $\sqrt{s}$ = 5.02, 7 and 13 TeV for forward and central multiplicity estimators.}\label{energy}
\end{figure}
PYTHIA8 predictions implemented with color reconnection (CR) in the string fragmentation process is compared with the multiplicity class (0.01-0.1\%) as shown in Fig.~\ref{norm_dndeta00101_forward}. This model, describes the data within 15$\%$, however, PYTHIA8 with no color reconnection overestimates the data by up to 20$\%$. We have also compared the PYTHIA6 Perugia tune and  EPOS-LHC models, which incorporates collective flow like effects in the simulation to the (0.01-0.1\%) multiplicity class  and they follow the same trend as PYTHIA8 with CR.  Figure~\ref{energy} shows the energy dependence of the pseudorapidity density distribution.
The evolution of the average  pseudorapidity density with center of mass energy for different multiplicity classes from low multiplicity to high multiplicity is parametrized by a power law function.
 The rise in the average  pseudorapidity density as a function of center-of-mass energy becomes steeper for high multiplicity classes, which could be due to multiple parton interactions \cite{mpi1,mpi2}.
 
\vspace{-7mm} 
\section{Strange hadron production}
\begin{figure}[h]
	\includegraphics[height=0.7\textheight,width=0.32\textwidth,keepaspectratio]{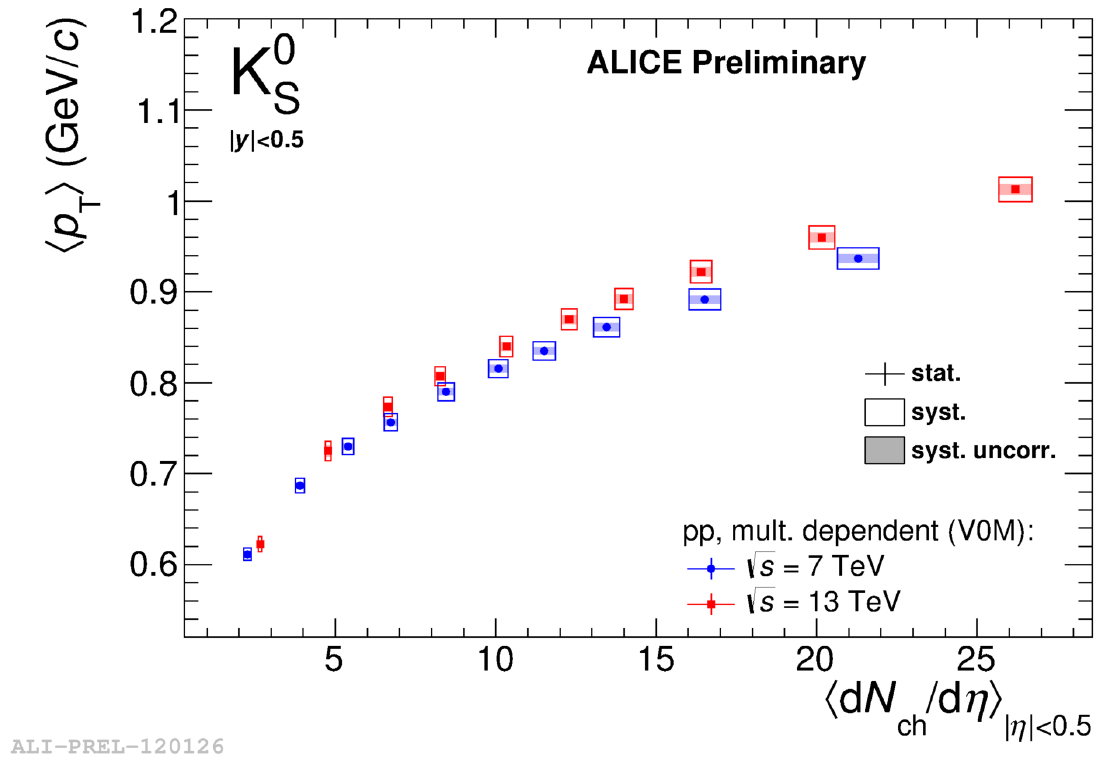}
	\includegraphics[height=0.7\textheight,width=0.32\textwidth,keepaspectratio]{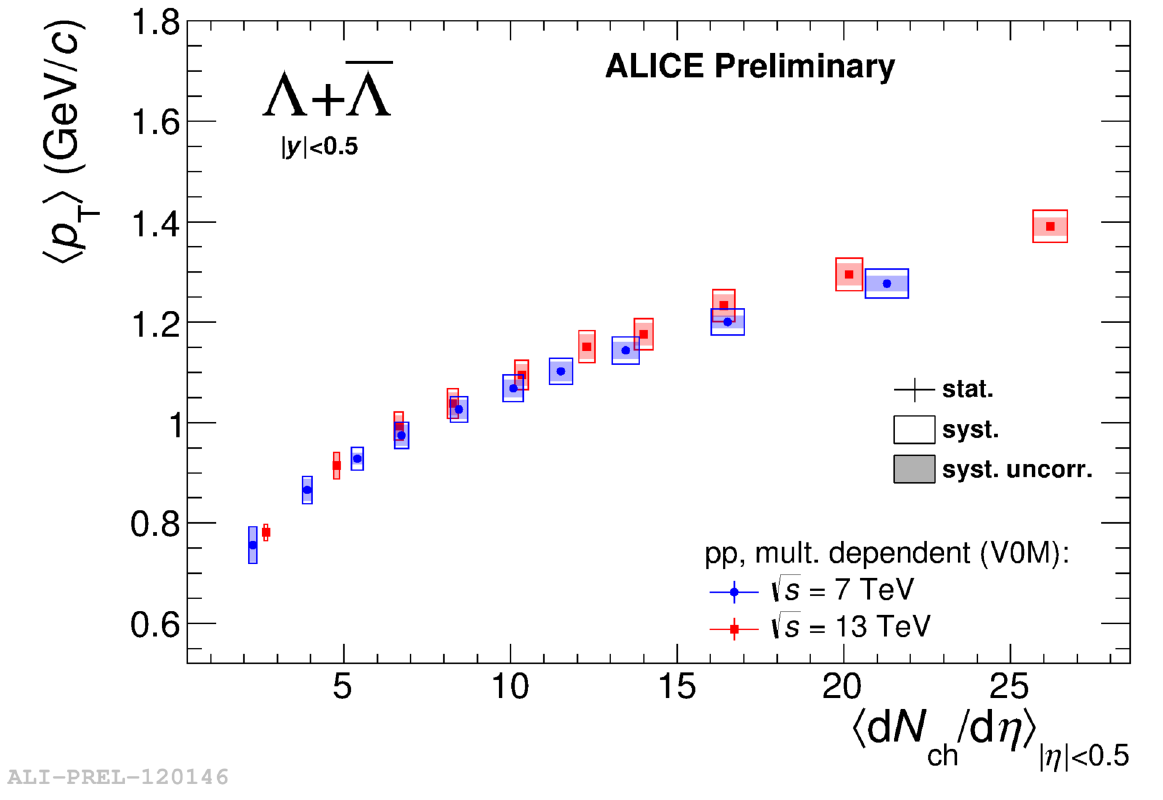}
	\includegraphics[height=0.7\textheight,width=0.32\textwidth,keepaspectratio]{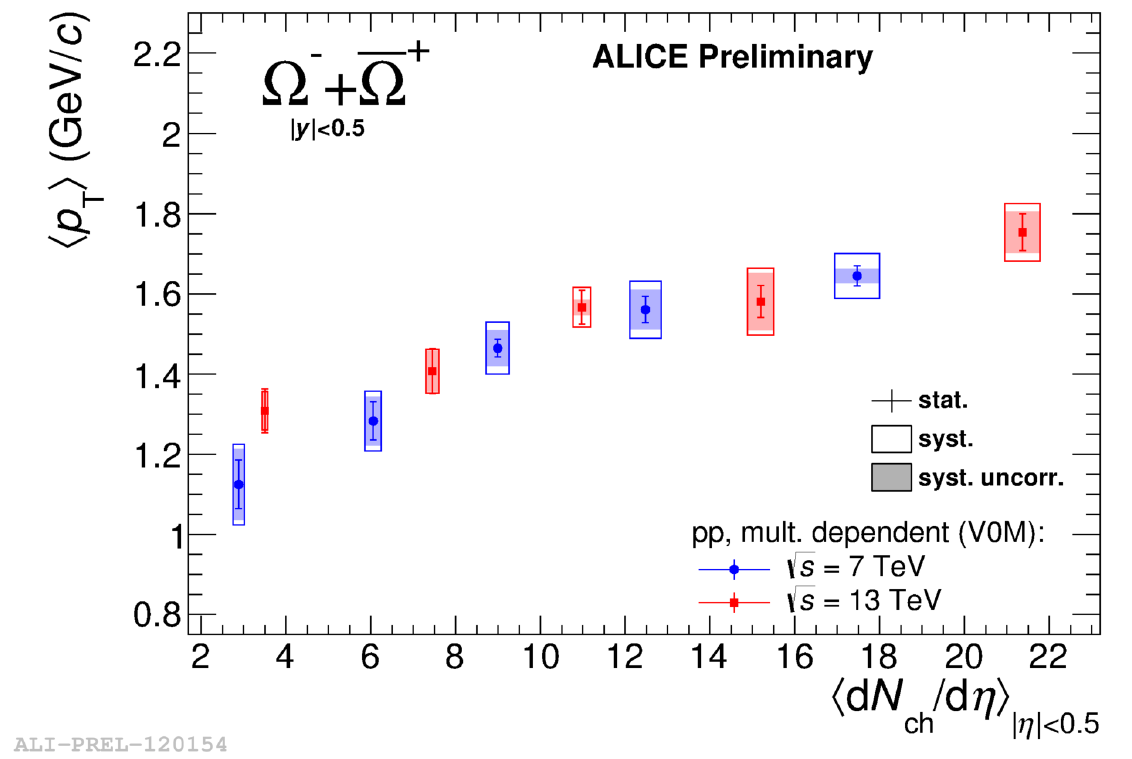}
\caption{ $\langle p_{T} \rangle$  of  $K_{S}^{0}$, $\Lambda$, and $\Omega$ as a function of multiplicity in forward multiplicity event classes at $\sqrt{s}$ =  7 and 13 TeV.}\label{strange2}
\end{figure}
New measurements of strangeness production in pp collisions at $\sqrt{s}$  = 13 TeV and in Xe-Xe collisions  at $\sqrt{s_{NN}}$  = 5.44 TeV  using new data samples will be discussed here. Figure~\ref{strange2} shows the energy dependence of the  $\langle p_{T} \rangle$  of strange hadrons versus the charged multiplicity at mid-rapidity, where results are compared to the previous pp measurements at
$\sqrt{s}$ =  7  TeV. The   $\langle p_{T} \rangle$  of $K_{S}^{0}$  as shown in Fig.~\ref{strange2} is marginally greater for 13 TeV  center-of-mass energy, where as for $\Lambda$ and $\Omega$ it scales with 13 TeV center-of-mass energy.
 \begin{figure}
	\centering
	\includegraphics[height=.9\textheight,width=0.36\textwidth,keepaspectratio]{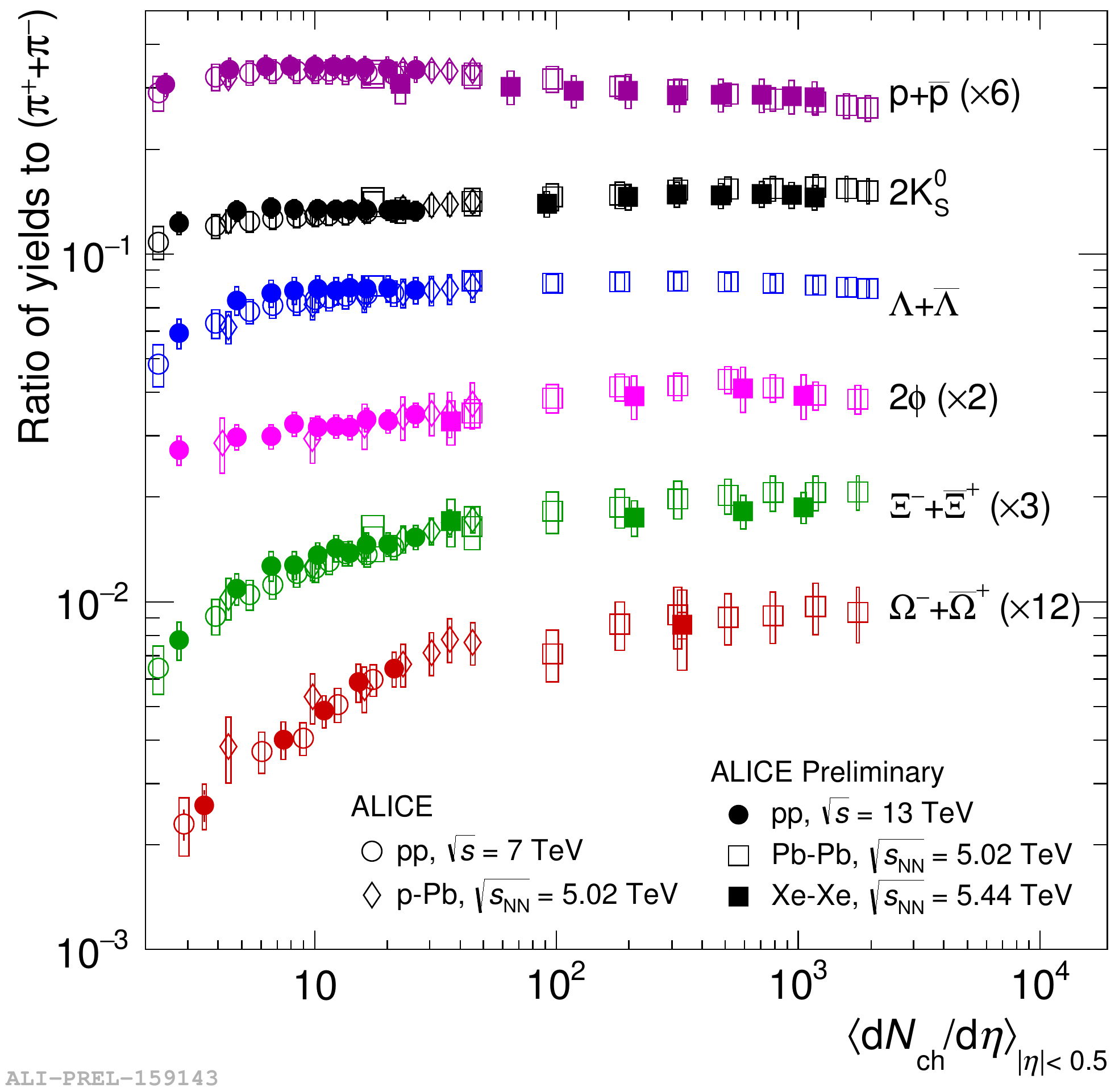}
	\caption{ Hadron-to-pion yield ratios as a function of the charged particle multiplicity in pp, p-Pb, Pb-Pb, and Xe-Xe collisions at the LHC.}\label{strange3}
\end{figure}
The evolution of the ratio of the total yield of strange particles to the yield of pions (in order to  quantify the strangeness enhancement)  as a function of charged particle multiplicity at mid-rapidity is shown in  Fig \ref{strange3}. This ratio is  measured for strange and non-strange hadrons including   p, $K_{S}^{0}$, $\Lambda$,  $\phi$, $\Xi$ and $\Omega$ and for all the available colliding systems. We observe a smooth evolution with multiplicity across all colliding systems including the Xe-Xe system, from low multiplicity pp to central A-A collisions. 
 
 \vspace{-0.1mm} 
\section{Conclusion}
To improve our understanding of various aspects of multiplicity dependent final state observable properties, in particular at high multiplicity, an unbiased measurement of $\langle \mathrm{d}\it{N}_{\mathrm{ch}}/\mathrm{d}\it{\eta} \rangle $ at mid-rapidity is very crucial.  
 
 In this contribution, we measured for the first time  the pseudorapidity distributions of charged particles in proton-proton collisions at $\sqrt{s}$ = 13 TeV for different multiplicity classes and estimators.  The average charged multiplicity density ($\langle \mathrm{d}\it{N}_{\mathrm{ch}}/\mathrm{d}\it{\eta} \rangle $)  increases steeply with center-of-mass energies for high multiplicity classes. Similar results and conclusions can be drawn for 5.02 and 7 TeV center-of-mass energies. We also report the first observation of multiplicity dependence of strangeness production in pp collisions at 13 TeV. These measurements show that strangeness yields scale well with the center-of-mass energy while the average momentum does not.
 
 \vspace{-3mm} 
\subsection{Acknowledgment}
This work was supported in part by Polish National Science Centre grant DEC-2016/23/B/ST2/01409, by the AGH UST statutory tasks No. 11.11.220.01/4 within subsidy of the Ministry of Science and Higher Education, and by PL-Grid Infrastructure.

\end{document}